\documentclass[aps,prl,reprint,showpacs,amsfonts,amsmath,floatfix,superscriptaddress,footinbib]{revtex4-2}

\usepackage[svgnames]{xcolor}

 \usepackage[colorlinks=true,        
            allcolors = black,  
            citecolor=DarkBlue, 
            linkcolor=DarkBlue,
            urlcolor=DarkBlue
        ]{hyperref}  
\usepackage{amsmath, amsfonts, amssymb, amsthm, bbm}
\usepackage{dsfont}
\usepackage{bm} 
\usepackage{mathtools}
\usepackage{times}
\usepackage{physics}
\usepackage{graphicx}
\usepackage{float}
\usepackage{enumitem}
\usepackage{dcolumn}
\usepackage{subfiles}
\theoremstyle{plain}
\newtheorem{theorem}{Theorem}

\newtheorem{lemma}{Lemma}

\newcommand{\ii}{\mathrm{i}}
\newcommand{\e}{\mathrm{e}}
\theoremstyle{definition}

\newcommand{\bal}{\begin{equation}\begin{aligned}}
\newcommand{\eal}{\end{aligned}\end{equation}}

\allowdisplaybreaks

\begin{document}
\title{Measuring non-Gaussianity with Correlation}

\author{Oliver Hahn}
\email{hahn@g.ecc.u-tokyo.ac.jp}
\affiliation{Department of Basic Science, The University of Tokyo, 3-8-1 Komaba, Meguro-ku, Tokyo, 153-8902, Japan}

\author{Ryuji Takagi}
\email{ryujitakagi.pat@gmail.com}
\affiliation{Department of Basic Science, The University of Tokyo, 3-8-1 Komaba, Meguro-ku, Tokyo, 153-8902, Japan}

\begin{abstract}
Quantum non-Gaussianity is a key resource for quantum advantage in continuous-variable systems. We introduce a general framework to quantify non-Gaussianity based on correlation generation: two copies of a state become correlated at a $50{:}50$ beam splitter if and only if the state is non-Gaussian, with correlations reducing to entanglement in the pure-state case. This connection enables operational measures of non-Gaussianity, defined through correlation quantifiers such as Rényi-$\alpha$ entropy for pure states and Rényi-$\alpha$ mutual information for mixed states. We prove that all such measures are monotonic under Gaussian channels. Building on this framework, we propose a sample-efficient experimental protocol to estimate non-Gaussianity using standard optical components, even in the state agnostic setting. Finally, we establish a lower bound on the sample complexity of estimating Wigner negativity, allowing a direct comparison with our protocol. Our results provide both a unifying theoretical framework for non-Gaussianity and a practical route toward its experimental quantification.

\end{abstract}

\maketitle

Quantum information processing exploits inherently quantum effects to achieve advantages over classical counterparts.
Among the various frameworks for implementing quantum protocols, continuous-variable (CV) systems, which are described by infinite-dimensional Hilbert spaces associated with observables possessing a continuous spectrum, are particularly promising.
This paradigm has been realized in various experimental platforms, including quantum optical systems~\cite{konno2024logical}, microwave cavities coupled to superconducting qubits~\cite{campagne2020quantum, sivak2023real}, and trapped ions~\cite{Fl_hmann_2019, de2022error}.

Gaussian states and operations play a central role in continuous-variable quantum information processing. Despite the infinite dimensionality of the underlying Hilbert space, Gaussian states can be fully characterized by their first and second moments --- namely, the mean vector and the covariance matrix. Moreover, their evolution under Gaussian operations is efficiently described at the level of these moments, enabling powerful analytical techniques.
Beyond their mathematical tractability, Gaussian states and operations are readily accessible in experimental platforms, particularly in quantum optics. Their experimental availability has enabled the implementation of various foundational quantum protocols, including quantum teleportation~\cite{PhysRevA.49.1473, PhysRevLett.80.869, PhysRevLett.81.5668} and quantum key distribution~\cite{PhysRevLett.88.057902}.

Restricting to Gaussian states and Gaussian operations imposes significant limitations. Within Gaussian quantum information, several no-go theorems prohibit key tasks such as entanglement and resource distillation~\cite{PhysRevLett.89.097901,PhysRevLett.89.137904,PhysRevA.66.032316,Lami2018Gaussian}, quantum error correction~\cite{PhysRevLett.102.120501}, and the violation of Bell inequalities and contextuality~\cite{PhysRevLett.129.230401}. Additionally, purely Gaussian processes can be efficiently simulated on classical computers~\cite{PhysRevLett.109.230503}, which precludes any computational advantage. Therefore, non-Gaussianity, whether in the form of states or operations, is essential for overcoming these limitations and enabling genuinely quantum protocols --- though realizing such resources experimentally remains a major challenge.

The operational distinction between Gaussian and non-Gaussian has motivated the development of a resource theory of non-Gaussianity. Resource theories aim to quantify the amount of a given resource --- in this case, non-Gaussianity --- required to implement specific tasks or protocols. Several quantifiers have been proposed, including the relative entropy of non-Gaussianity~\cite{PhysRevA.78.060303,PhysRevA.82.052341,PhysRevA.88.012322}, robustness~\cite{Regula2021operational,Lami2021framework,turner2024nongaussianstatesadvantageouschannel}, stellar rank~\cite{PhysRevLett.124.063605,hahn2024classicalsimulationquantumresource}, Gaussian rank or extent~\cite{hahn2024classicalsimulationquantumresource, PhysRevA.110.042402} and Wigner negativity~\cite{Kenfack2004negativity,PhysRevA.97.062337,PhysRevA.98.052350}.

These measures not only serve as formal tools within the resource-theoretic framework but also allow one to quantify how non-Gaussian—and thus how genuinely quantum—a given experiment is.
However, a major limitation of many existing measures is that their evaluation typically requires full quantum state tomography, which becomes impractical as the number of modes increases~\cite{mele2024learningquantumstatescontinuous}.

\begin{figure}
    \centering
    \includegraphics[width=0.45\textwidth]{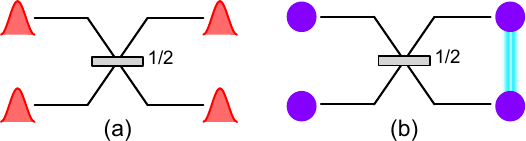}  
    \caption{The underlying idea of our framework.
    (a) The output of two copies of the same Gaussian state  though a $50{:}50$ beam splitter results in a product state in the output. (b) Conversely two copies of the same non-Gaussian state such as Fock states will be correlated after the $50{:}50$ beam splitter. In consequence we use the correlation generated after the beam splitter to detect and quantify the non-Gaussianity of the input state. For pure states the correlation is equivalent to the entanglement in the output.}
    \label{fig:construction}
\end{figure}

In this work, we introduce a novel framework for defining measures of non-Gaussianity, which allow in some cases efficient experimental access, based on the generation of correlation~\cite{Henderson_2001}. Specifically, when two identical copies of a state are sent through a $50{:}50$ beam splitter, the output state is correlated if and only if the input state is non-Gaussian~\cite{10.1063/1.5122955} (see Figure~\ref{fig:construction}). 
This observation enables us to use measures of correlation applied to the output of the beam splitter as operational quantifiers of non-Gaussianity. We establish general monotonicity properties for this class of measures, ensuring their validity within a resource-theoretic framework.

We focus in particular on two such measures: the Rényi-$\alpha$ entropy and Rényi-$\alpha$ mutual information. The Rényi-$2$ entropy, in particular, is experimentally attractive, as it can be efficiently estimated using a SWAP test, even without prior knowledge of the quantum state --- making the protocol state-agnostic.
Additionally, we derive a lower bound on the sample complexity required to estimate the Wigner negativity and compare it to the sampling cost of estimating the Rényi-2 entropy. Our analysis reveals that the latter can be significantly more efficient.

Overall, our results provide a new operational interpretation of non-Gaussianity through its ability to generate correlation. This connection not only provides an operational interpretation of non-Gaussian resources but also offers practical tools for experimental quantification. Moreover, our approach can be viewed as a generalization of the Hong–Ou–Mandel experiment to arbitrary quantum states and correlation measures.

\paragraph*{Background. ---}
We consider an infinite-dimensional separable Hilbert space consisting of $m$ modes.
The canonical operators or quadratures are $\hat{\bm{r}}=(\hat q_1,\hat p_1,...,\hat q_m,\hat p_m)^T$ with commutation relation
\begin{align}
    \qty[\hat{\bm{r}},\hat{\bm{r}}^T]=\ii\Omega
\end{align}
with 
$\Omega = \bigoplus^{m}_{j=1} \begin{pmatrix}
    0& 1\\
    -1 &0 
    \end{pmatrix}$ being the symplectic form.
Note that $\hat{\bm{r}} \hat{\bm{r}}^T$ is the outer product.
It is often convenient to represent quantum states using the characteristic function $\chi_\rho(\bm{r})=\Tr\qty[\hat{D}(\bm{r})\rho]$, where $\hat{D}(\bm{r})=\e^{\ii\bm{r}\Omega \hat{\bm{r}}}$ is the displacement operator. 
This allows us to define the set of Gaussian states in an instructive manner.
We define the set of Gaussian states $\rho_G \in \mathcal{G}$ as the states that cen be represented as
\begin{align}
    \chi_{\rho_G}(\bm{r})=\e^{\frac{1}{4}\bm{r}^T\Omega^T Y \Omega \bm{r}-\ii \bm{\mu}\Omega \bm{r}}.
\end{align}
Thus the characteristic function of a Gaussian state is represented by a multivariate Gaussian function.
Gaussian unitaries map Gaussian pure states to Gaussian pure states and are generated by Hamiltonians that are at most quadratic in the quadrature operators.
The Gaussian unitary that is central to this work is the $m-$mode $50:50$ beam splitter
\begin{align}
    \hat{U}_{BS}=\exp\qty(\ii \frac{\pi}{4}\sum_{i=1}^m  \left[\hat{p}_{i,A} \hat{q}_{i,B} -\hat{q}_{i,A} \hat{p}_{i,B} \right]).
\end{align}
Every Gaussian unitary $\hat{U}_G$ has an associated symplectic matrix $S_G$ that transforms the vector of quadratures as $\hat{U}_G \hat{ \bm{r}}\hat{U}_G^\dagger = S_G \hat{\bm{r}}$.
The symplectic matrix of the beam splitting unitary defined above is
\begin{align}
        S_{BS}=\frac{1}{\sqrt{2}}\begin{pmatrix}
        \mathds{1}_m&\mathds{1}_m\\
        -\mathds{1}_m&\mathds{1}_m
    \end{pmatrix}
\end{align}
where the ordering is $\bm{r}=[q_1,p_1,...,q_m,p_m]^T$.
For more details, see Appendix~\ref{ap:background}.

Gaussian channels are completely-positive and trace-preserving maps that map any Gaussian states to another Gaussian states.
We can write any Gaussian channel from $n$ modes to $m$ modes using a Stinespring dilation~\cite{serafini2023quantum} using vacuum states and Gaussian unitaries as
\begin{align}
    \mathcal{E}_G(\rho)=\Tr_C\qty[\hat{U}_G \rho\otimes \dyad{0}^{\otimes{k}}\hat{U}_G],
\end{align}
where the subsystem $C$ contains the last $n+k-m$ modes.

\paragraph*{Correlation through beam splitting. ---}
We start with the central construction of this manuscript.
We consider two copies of the same state and act on these states with an $m-$mode $50{:}50$- beam splitter
\begin{align}
    \chi_{\hat U_{BS}\rho\otimes \rho\hat U_{BS}^\dagger}\qty(\bm{r}=\qty(\bm{r_1},\bm{r_2})^T)=\chi_{\rho}\qty(\frac{\bm{r_1}-\bm{r_2}}{\sqrt{2}}) \chi_{\rho}\qty(\frac{\bm{r_2}+\bm{r_1}}{\sqrt{2}}).
\end{align}
Importantly, the output state of the beam splitter is a product state if and only if the input state is Gaussian~\cite{10.1063/1.5122955}.
This implies that the state $\rho_{AB}= \hat{U}_{BS} \rho\otimes \rho\hat{U}_{BS}^\dagger$ is correlated if and only if $\rho$ is a non-Gaussian state.
Moreover, in the case of pure states, $\Psi_{AB}= \hat{U}_{BS} \psi\otimes \psi\hat{U}_{BS}^\dagger$ is entangled if and only if $\psi$ is a pure non-Gaussian state.

Having the direct connection between non-Gaussianity and correlation generation established, we 
can define our framework to define measures of non-Gaussianity.
Let $C$ be an arbitrary measure of correlation and define $N_C(\rho)\coloneqq C(\hat U_{BS}\rho\otimes\rho \hat U_{BS}^\dagger)$. 
By identifying the class of operations on the input state copies that cannot increase the resulting correlations, we can derive general structural properties of our construction. 
The resulting monotone $N_C$ is then a faithfull monotone under Gaussian channels.
\begin{theorem}[Properties]
    The measure of non-Gaussianity $N_C(\rho)\coloneqq C(\hat U_{BS}\rho\otimes\rho \hat U_{BS}^\dagger)$ with $C$ being an arbitrary measure of correlation is monotonic under
    Gaussian channels. Furthermore, $N_C(\rho)$ is a faithful measure, i.e. $N_C(\rho)=0$ if and only if $\rho$ is a Gaussian state.
\end{theorem}
The proof can be found in Appendix~\ref{ap:monotonicity}.

Another general property is that if the input state is a product state, $\rho = \rho_A \otimes \rho_B$, then the output of the beam splitter remains factorized., i.e. $\hat{U}_{BS} \rho \otimes \rho \hat{U}_{BS}^\dagger = \qty[\hat{U}_{BS} \rho_A \otimes \rho_A\hat{U}_{BS}^\dagger ]\otimes \qty[\hat{U}_{BS} \rho_B \otimes \rho_B\hat{U}_{BS}^\dagger ]$. This is a useful property if the measure $E$ is multiplicative or additive. Otherwise, this property can be useful to derive bounds.

As mentioned above $N_C$ is a faithful measures of non-Gaussianity.
Considering the fact that preparing the probabilistic mixture of Gaussian states is operationally easily accessible, another concept known as genuine non-Gaussianity has been discussed~\cite{PhysRevA.97.062337,PhysRevA.98.052350}. 
Since the set of Gaussian states is non-convex, a probabilistic mixture of two Gaussian states can be non-Gaussian. 
A state is called genuinely non-Gaussian if it cannot be written as a convex combination of Gaussian states. 
The natural question in the context of this work is whether our measure serves as a measure of genuine non-Gaussianity. 
However, as stated before the output is a product state if and only if the input states are Gaussians, which implies that the output is generally correlated.
Thus it is not the case by showing that our measure generally takes a non-zero value for a convex mixture of Gaussian states, clarifying that our measure is a valid measure for the resource theory of non-Gaussianity that takes the set of (non-convex) Gaussian states as its free states as in Refs.~\cite{PhysRevA.78.060303,PhysRevA.82.052341}.

With the general construction and their properties in place, we can now focus on specific measures of correlation to quantify the non-Gaussianity of the input states.
For pure states the correlation of the state and the entanglement of the state are equivalent.
Therefore we consider the entanglement Rényi-$\alpha$ entropy~\cite{PhysRevA.93.022324}
\begin{align}
    E_{\alpha}(\rho)=\frac{1}{1-\alpha}\log_2\qty(\Tr\qty[\Tr_B\qty(\rho)^\alpha])
\end{align}
for the pure state case.
For $\alpha=1$, this reduces to the von Neumann entropy of the reduced state, and for $\alpha=2$ it is directly related to the purity of the reduced state. The Rényi-$\alpha$ entropy serves as an entanglement measure for pure states and is particularly straightforward to compute for $\alpha=2$, as it requires only the reduced-state purity --- a quantity that, as we will show later, can be accessed efficiently in experiments.

The direct connection between purity of the reduced state and the non-Gaussianity allows us to find a requirement on the characteristic function.
Note that purity is a quantity involving the square of the characteristic function, i.e. 
$\Tr\qty[\rho^2]=\frac{1}{(2\pi)^m}\int \dd \bm{r} \abs{\chi_{\rho}(\bm{r})}^2 $, while non-Gaussianity involves the quartic power of the characteristic function
\begin{align}
    \Tr\qty[\qty( \Tr_B\qty[ \hat{U}_{BS} \rho \otimes \rho \hat{U}_{BS}^\dagger ] )^2 ]=\frac{1}{(2\pi)^m} \int \dd \bm{r} \abs{\chi_{\rho}\qty(\frac{\bm{r}}{\sqrt{2}})}^4.
\end{align}

The measures discussed so far are valid only for pure states, where correlations and entanglement are equivalent. For mixed states, however, one must consider all correlations, not just entanglement, since there exist non-Gaussian states that remain separable but are no longer product after the beam splitter. Further details are provided in Appendix~\ref{ap:corrnotent}.
In order to treat mixed states as well, we consider the Rényi-$\alpha$ mutual information~\cite{10.1063/1.5143862}
\begin{align}
    I_{\alpha}(\rho_{AB})=\min_{\sigma_B} D_{\alpha}\qty(\rho_{AB}\lVert \rho_A\otimes \sigma_B),
\end{align}
where $D_{\alpha}\qty(\cdot \lVert \cdot )$ is the minimal Rényi divergence with $\alpha \in [\frac{1}{2},\infty)$ and $\rho_A=\Tr_B[\rho_{AB}]$.
For $\alpha=1$ one recovers the known von Neumann mutual information
\begin{align}
    I_1(A;B)_\rho&=D(\rho_{AB}\lVert \rho_A \otimes \rho_B)
\end{align}
where $D\qty(\cdot \lVert \cdot )$ is the Umegaki's relative entropy.
Evaluating the Rényi-$\alpha$ mutual information for arbitrary $\alpha$ is challenging as it involves a minimization over states. However, there are upper and lower bounds using Rényi-$\alpha$ entropy and Rényi-$\alpha$ conditional entropy~\cite{10.1063/1.5143862}, which could admit direct estimation. Consult Ref.~\cite{Tomamichel_2016} for more information on 
Rényi-$\alpha$ (conditional) entropy.

\begin{figure}
    \centering
    \includegraphics[width=0.45\textwidth]{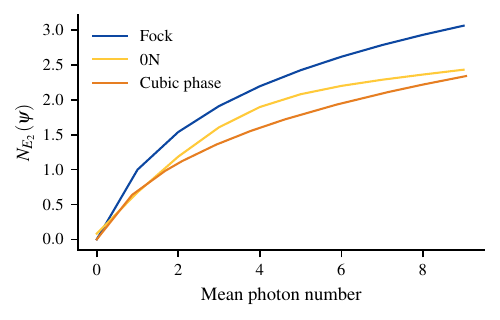}  
    \caption{This figures shows the measure of non-Gaussianity $N_{E_2}$ using the
    entanglement $2$-Rényi entropy depending on the mean photon number of a Fock state, a $0N$ state and a cubic phase state. We see that the Fock states has the highest value for the same mean photon number. }
    \label{fig:E2}
\end{figure}

In Appendix~\ref{ap:examples}, we present analytical examples for a variety of states, evaluating the Rényi-$\alpha$ entropy. 
In Figure~\ref{fig:E2} we present the entanglement Rényi-$2$ entropy for various states with respect to the mean photon number.
Interestingly, we observe that different classes of standard non-Gaussian states take different values, where Fock states achieve the largest.
This contrasts with the case of Wigner negativity, where these states were found to have nearly the same values for fixed mean photon number~\cite{PhysRevA.97.062337}.
This means that our measure can provide further restrictions on feasible transformation between different classes of non-Gaussian resource states that could not be detected by the Wigner negativity.

\paragraph*{Experimental access. ---}
Having established the theoretical framework, we now turn to the practical question of how these quantities can be measured in an experimental setting.

As noted previously, Gaussian states are amenable to many analytical tools because they are fully characterized by their covariance matrix and mean vector. However, every quantum state—including non-Gaussian ones—has an associated covariance matrix, even though it does not fully determine the state in the non-Gaussian case. A natural first step experimentally is to measure the covariance matrix and apply established methods to infer partial information about correlations. However, the correlations generated in our setting cannot be captured by the covariance matrix alone.
\begin{lemma}
The covariance matrix of the two input copies before and after the $50{:}50$ beam splitter is identical.
\end{lemma}
Consult Appendix~\ref{ap:not_in_cov} for the proof.

Since methods relying solely on the covariance matrix fail to detect the correlation generated in our approach, we turn to a promising alternative. As demonstrated earlier, the Rényi-$\alpha$ entropy is an effective tool to detect and quantify non-Gaussianity for pure states as then entanglement and correlation are equivalent. Equivalently, for $\alpha=2$ this corresponds to measuring the purity of one of the beam splitter outputs.
This approach is closely related to convolution operations in continuous-variable systems~\cite{Becker_2021} and finds analogues in recent works on quantifying non-Gaussianity in fermionic systems~\cite{coffman2025measuringnongaussianmagicfermions,lyu2024fermionicgaussiantestingnongaussian} and qudits~\cite{Bu_2023,bu2023stabilizertestingmagicentropy,Bu2025quantum}.

It is well known that the purity of an unknown quantum state can be determined experimentally via a SWAP test~\cite{PhysRevLett.88.217901}, a ubiquitous tool in quantum information processing. The standard implementation involves a controlled-SWAP gate acting on two copies of the state, conditioned on the state of an auxiliary qubit. The purity is then obtained by repeated measurement on the auxiliary qubit.
In consequence, measuring non-Gaussianity involves generating two copies of the state after the beam splitter and determining the purity of one subsystem. This procedure requires four copies of the original state we want to test.

\begin{figure}
    \centering
    \includegraphics[width=0.4\textwidth]{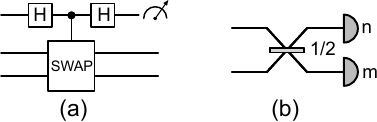}  
    \caption{Circuits for experimentally measuring the purity. (a) Measuring routine for a standard SWAP test. A qubit acts as the control of the control SWAP gate on the CV-modes. If the CV modes are two copies of the same state $\rho$, the measurement statistics of the qubit will be equivalent to the purity of the state $\rho$. (b) Circuit of a CV destructive SWAP test. Both modes are measured using PNR detectors after a $50{:}50$ beam splitter. In principle only measuring one output mode is necessary, however measuring two can deal with saturation in the detectors~\cite{Volkoff2022ancillafree}.}
    \label{fig:circuits}
\end{figure}

A SWAP test using a controlled-SWAP gate has been experimentally realized in a microwave architecture through coupling with a transmon qubit~\cite{Gao_2019}. Building on this, a recent proposal aims to improving the error resilience of the experimental protocol~\cite{Pietikäinen_2022}.
Rather than implementing a controlled-SWAP gate directly, one can apply beam splitters to the two CV states, combined with a controlled phase-shift between one of the states and a coupled qubit~\cite{PhysRevA.65.062320}. The purity of the state can then be extracted through repeated measurements of the coupled qubit. 
All of these implementations of a SWAP-test of CV-modes require coupling to an auxiliary qubit.
Coupling between qubits and continuous-variable modes is common in microwave architectures, making this approach particularly well-suited for such platforms.
The circuit of a standard SWAP test is depicted in Figure~\hyperref[fig:circuits]{~\ref*{fig:circuits}a.)}.

The protocols for SWAP-test discussed so far are not easy to implement in quantum optical experiments, as coupling between CV nodes and qubits remains challenging.
However, in some specific cases such as restricting to Fock states controlled-SWAP gates have been implemented using photonic circuits~\cite{ono2017implementation}.
More generally, one can implement a destructive SWAP-test using photon number resolving (PNR) detectors~\cite{PhysRevA.87.052330}.
A destructive SWAP-test using this approach and Fock states as the input is equivalent to the seminal Hong-Ou-Mandel experiment~\cite{PhysRevA.87.052330}.
More recently a generalization of the Hong-Ou-Mandel experiment has been implemented experimentally using parity measurements to obtain the purity between different Fock states
~\cite{Islam_2015}.

While destructive SWAP tests discussed so far have been restricted to Fock states, it has been shown that a destructive SWAP test can be implemented for arbitrary states using a balanced beam splitter and two photon-number-resolving detectors~\cite{Volkoff2022ancillafree}.
This approach also works in the multi-mode scenario and was demonstrated experimentally~\cite{Volkoff2022ancillafree}.
A circuit of the destructive SWAP test can be found in Figure~\hyperref[fig:circuits]{~\ref*{fig:circuits}b.)}.
Realistic PNR detectors saturate beyond a finite photon number. We model this limitation by assuming that the detectors can resolve photon numbers up to a total of $m+n \leq 2M$, where $n,m$ are the measured photons of each detector respectively. Without this restriction, only a single PNR detector would be required. Under this assumption, the estimator for the SWAP test restricted to the $2M$-photon Fock subspace takes the form
$\Tr\qty[\text{SWAP}_{2M}\rho\otimes\sigma]=\frac{1}{S}\sum_{s=1}^S (-1)^{n(s)}\Theta\qty(2M-n(s)-m(s))$,  where $\Theta$ is the Heaviside function and $n(s),m(s)$ the measurement result of the PNR detectors at round $s$. 
The systematic error given by the finite detection capability of the PNR detectors can be bounded by
\begin{align}
    \abs{\Tr\qty[\rho \sigma] - \Tr\qty[\text{SWAP}_{2M} \rho\otimes \sigma]} \leq 1-q_{2M} \leq 1- q_M^\rho q_M^\sigma
\end{align}
with $q_M^{\rho/\sigma} =\Tr\qty[ \rho/\sigma \sum_{n=0}^M\dyad{n} ]$and $q_{2M}=\Tr\qty[Q_{2M} \rho \otimes \sigma] $ while $Q_{2M}$ is the projector onto the $2M$ Fock state subspace on $2$ modes.

In addition to this systematic error incurred through the PNR detector, we have a statistical error given by finite measurement statistic. 
 We can estimate the expectation value with standard deviation $\delta$ by measuring $\qty(\frac{1}{\delta^2})$ number of times.
The total error is upper bounded by the sum of both errors.
Thus in order to bound the total error by $\epsilon$, we require the total measurable photon number $2M$ has to fulfill $1- q_{2M}\leq  \epsilon$. This is a requirement on the quality of the PNR detectors.
We then have to run the  algorithm $\qty(\frac{1}{\delta^2})$ times where $0<\delta \leq \epsilon -\qty(1-  q_{2M})$.
The number of samples to obtain error $\epsilon$ therefore depends indirectly on the energy of the states one considers as the error depends how the tail in the Fock state decomposition decays. 

In all the cases discussed so far, the number of measurements required ---while ignoring systematic errors--- to estimate the purity with error $\epsilon$ scales as $\mathcal{O}\qty(\frac{1}{\epsilon^2})$, similar to the cost of estimating a single observable. In the standard SWAP test using qubit measurements, there are only two possible measurement outcomes. In contrast, PNR detectors theoretically yield a countably infinite set of outcomes, which might suggest an energy-dependent scaling. However, the key advantage of using PNR detectors lies in the fact that the purity depends solely on the parity of the detector’s outcome, simplifying the estimation considerably.

Overall, all methods proposed here to measure the non-Gaussianity of an arbitrary and unknown pure state requires four copies at the time, constant depth and $\mathcal{O}\qty(\frac{1}{\epsilon^2})$ rounds. This protocol allows us to measure non-Gaussianity efficiently and state-agnostically in an experiment. The advantage becomes clear when contrasted with full state tomography, which for pure states scales as $\mathcal{O}(\frac{E^m}{\epsilon^{2m}})$~\cite{mele2024learningquantumstatescontinuous}, where $E$ is the energy of the state. 

The mixed-state case presents additional challenges. In particular, an experimental determination of the Rényi-$\alpha$ mutual information is demanding, and we leave the design of a concrete protocol to future work. Nevertheless, a few general remarks are in order.
Even though Rényi-$\alpha$ mutual information includes a minimization we can use the bounds provided in~\cite{10.1063/1.5143862} to find bounds on the resource content.
Then, one can use the expression for the conditional Rényi-$\alpha$ entropies provided in~\cite{Tomamichel_2016} to evaluate the bounds, which could be measured using, e.g., the quantum singular value transformation~\cite{10.1145/3313276.3316366}, while thorough investigation on a specific protocol is left for further work.
See Appendix~\ref{ap:meas_mutual} for more details.

To conclude this paragraph, we note that if one cannot have access to several copies simultaneously, one can use classical shadow techniques~\cite{Becker_2024}. See Appendix~\ref{ap:shadows} for details.

\paragraph*{Lower bound on estimating Wigner negativity. ---}
In the previous section, we have seen that the amount of non-Gaussianity for pure states can be measured efficiently in experiments --- in principle independent of the number of modes or the energy. To assess the practical relevance of our proposal, we compare its experimental attainability with that of estimating the Wigner negativity.

\begin{figure}
    \centering
    \includegraphics[width=0.45\textwidth]{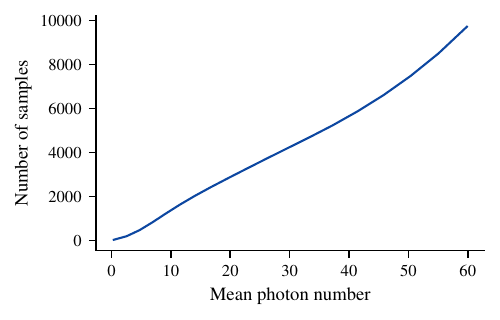}  
    \caption{Lower bound of the sample complexity to estimate the Wigner negativity of a state. The value is obtained by computing exact values of Wigner negativity for the cubic phase state.} 
    \label{fig:lower_bound}
\end{figure}

Wigner negativity~\cite{Kenfack2004negativity,PhysRevA.97.062337,PhysRevA.98.052350} is one of the mostly studied measures for the resource theory of non-Gaussianity and is widely used in experiments to show non-classicality of experimental states~\cite{PhysRevLett.87.050402,science.1122858}. Wigner negativity is a standard benchmark for assessing the quality of experimentally generated states. In practice, it is typically inferred via tomography adapted to the experimental platform, with the negativity then computed from the reconstructed state. However, this is an unfair comparison, as a full state reconstruction contains far more information than a single Wigner negativity value. Consequently, directly estimating the Wigner negativity—without reconstructing the entire state—could in principle be more efficient than state tomography.
To level the playing field, we establish a lower bound on the sample complexity required to estimate the Wigner negativity. Consult Appendix~\ref{ap:lower_bound} for details.
The basic idea of the lower bound is to connect the difference in Wigner negativity of two states with a state discrimination task.
Aside from the general connection between estimating Wigner negativity and state discrimination, we use a family of cubic phase states to quantitatively evaluate the lower bound.
Using a lower bound on the Wigner negativity of the cubic phase state, we show that the sample complexity diverges if the energy is not taken into account. Restricting to finite-mean photon number states renders the sample lower bound finite. In Figure~\ref{fig:lower_bound}, we illustrate the scaling of the lower bound on estimating the Wigner negativity as a function of the mean photon number.
Estimating the Wigner negativity therefore requires a sample complexity that scales at least as the third root of the mean photon number. Comparing this with the lower bound for state tomography, $\mathcal{O}\left(\frac{E}{\epsilon^{2}}\right)$ for a single mode, we see a gap in which a protocol that directly estimates the Wigner negativity could offer an advantage.
Nonetheless, this shows the efficiency of our approach to detect and measure non-Gaussianity in an experiment, since the number of samples required to measure non-Gaussianity using our protocols requires constant number of samples independent of the energy.

\paragraph*{Discussions. ---}
In summary, we have introduced a novel framework to detect and quantify non-Gaussianity in continuous-variable systems by linking it to the correlation generated when two copies of a state are mixed on a $50{:}50$ beam splitter. This fundamental connection allows us to define non-Gaussianity measures through the correlation produced at the beam splitter, extending earlier ideas applied to non-classical states~\cite{PhysRevA.65.032323,PhysRevLett.94.173602}. We have established general properties of this construction such as monotonicty under Gaussian channels and illustrated it using the Rényi-$\alpha$ entropy and the Rényi-$\alpha$ mutual information.
Experimentally, our framework enables the use of SWAP tests to measure non-Gaussianity with constant sample complexity representing a generalization of the seminal Hong–Ou–Mandel experiment. We also derived a lower bound on the number of sample required to estimate the Wigner negativity, showing the advantage of our proposed protocol.

Looking ahead, it will be valuable to develop continuous-variable–specific methods for directly measuring mutual information in experiments. 
Another promising direction is to explore the topology of non-Gaussian states within our framework, identifying which states are most resourceful for a given energy or which non-Gaussian states lead to genuinely entangled states and which to separable states. 
Finally, our sample lower bound on estimating Wigner negativity motivates the design of experimental protocols to estimate it directly—potentially without full tomography—or to establish a formal equivalence between direct estimation and full state reconstruction.

 \let\oldaddcontentsline\addcontentsline
\renewcommand{\addcontentsline}[3]{}

\begin{acknowledgments}
This work is supported by JST CREST Grant Number JPMJCR23I3, JSPS KAKENHI Grant Number JP24K16975, JP25K00924, and MEXT KAKENHI Grant-in-Aid for Transformative
Research Areas A ``Extreme Universe” Grant Number JP24H00943.
 \end{acknowledgments}
\paragraph*{Note added.}
 During the preparation of this manuscript, we became aware of the independent work of Bu and Li~\cite{bu2025efficientmeasurementbosonicnongaussianity}.

 \bibliography{bib}


\clearpage
\onecolumngrid
\let\addcontentsline\oldaddcontentsline
\appendix

 \renewcommand{\tocname}{Supplemental Material}

\tableofcontents

\makeatletter
\let\toc@pre\relax
\let\toc@post\relax
\makeatother 

\section{Background: Continuous variables}
\label{ap:background}

We start the supplemental material by introducing the necessary background for continuous-variable quantum systems~\cite{serafini2023quantum} .
In this work, we will use the canonical operators $\hat q,\hat p$ that fulfill the following commutation relations
\begin{align}
    [\hat q,\hat p]=\ii.
\end{align}
The commutation relations can be conveniently written for a vector of quadratures of $m-$modes $\hat{\bm{r}}=(\hat q_1,\hat p_1,...,\hat q_m,\hat p_m)^T$ as
\begin{align}
    \qty[\hat{\bm{r}},\hat{\bm{r}}^T]=\ii\Omega
\end{align}
with 
$\Omega = \bigoplus^{m}_{j=1} \begin{pmatrix}
    0& 1\\
    -1 &0 
    \end{pmatrix}$ being the symplectic form.
Note that $\hat{\bm{r}} \hat{\bm{r}}^T$ is the outer product.

Gaussian states and Gaussian unitaries play a central role for continuous variables.
Gaussian unitaries transform the canonical operators linearly are defined as
\begin{align}
    \hat U_G&=e^{\ii \hat H}\\
    \hat H&= \frac{1}{2}\hat{\bm{r}}^T\text{H}\hat{\bm{r}}+\bm{\Bar{r}}\hat{\bm{r}}
\end{align}
where we used the notation $\bm{\Bar{r}}= (r_{q_1},r_{p_1},...r_{q_m},r_{p_m})^T$ a vector of real numbers.
The matrix $\text{H}$ is a symmetric matrix, while $\hat{H}$ denotes the operator.
Every quadratic Gaussian unitary operator has an associated symplectic matrix
\begin{align}
    S= \e^{\Omega \text{H}}
\end{align}
fullfilling
\begin{align}
    \hat U_G \hat{\bm{r}}  \hat U_G^\dagger = S \hat{\bm{r}}.
\end{align}
Symplectic matrices $S\in \text{Sp}_{2m}(\mathds{R})$ are defined as the matrices that keep the symplectic form invariant
\begin{align}
    S\Omega S^T=\Omega.
\end{align}

The operators generated by the Hamiltonian of linear order are the so-called Heisenberg-Weyl operator or Displacement operator
\begin{align}
    \hat{D}(\bm{\Bar{r}})=\e^{\ii \bm{\Bar{r}}^T\Omega \hat{\bm{r}}}
\end{align}
that transform the canonical operators like
\begin{align}
   \hat{D}(\bm{\Bar{r}}) \hat{\bm{r}}\hat{D}(\bm{\Bar{r}})^\dagger =  \hat{\bm{r}}+\bm{\Bar{r}}.
\end{align}
Any (multimode) Gaussian unitary can, as a consequence of the Euler (sometimes called Bloch--Messiah) decomposition, be decomposed as
\begin{align}
    \hat U_G= \hat{U} \hat{S}(s) \hat{D}(\bm{r})   \hat{V},
\end{align}
where $\hat{U}$ and $\hat{V}$ are passive Gaussian unitary operators.
The unitary $\hat{S}(s)=\bigotimes_{i=1}^m  \hat{S}(s_i) $ is called squeezing
\begin{align}
    \hat{S}(s_i)=\e^{\ii \frac{s_i}{2}\qty(\hat{q}\hat{p} + \hat{p}\hat{q})}.
\end{align}
Passive Gaussian unitaries are energy-preserving and can in turn be decomposed into a product of phase rotations 
\begin{align}
    \hat{U}_R\qty(\phi)=\e^{\ii\frac{\phi}{2}\qty(\hat{q}^2+\hat{p}^2)}
\end{align}
and beam splitter 
\begin{align}
    \hat{U}_{BS}(\theta)=\exp\qty(\ii \theta\sum_{i=1}^m  \left[\hat{p}_{i,A} \hat{q}_{i,B} -\hat{q}_{i,A} \hat{p}_{i,B}\right] )
\end{align}
respectively. 
The unitary implements for $\theta = \frac{\pi}{4}$ a $50{:}50$-beam splitter. In the following we will denote $\hat{U}_{BS}$ being the $50{:}50$-beam splitter.

Analogously, we define Gaussian states as the states that are generated by quadratic Hamiltonian
\begin{align}
    \rho_G=\frac{e^{-\beta \hat H}}{\Tr[e^{-\beta \hat H}]}
\end{align}
including the case $\beta\rightarrow\infty$ being pure states.
Gaussian states have the important property that they are fully determined by their mean
\begin{align}
    \bm{\Bar{r}}=\Tr[\hat{\bm{r}}\rho]
\end{align}
and their covariance matrix
\begin{align}
    \sigma= \Tr[\{(\hat{\bm{r}}-\bm{\Bar{r}}),(\hat{\bm{r}}-\bm{\Bar{r}})^T\}\rho]
\end{align}
where $\{,\}$ is the anti-commutator. 
Note that although Gaussian states are fully determined by their first two statistical moments, higher-order moments generally do not vanish. Non-Gaussian states also possess a well-defined mean and covariance matrix, but they are not fully characterized by these quantities.

Gaussian states can be conveniently be represented the characteristic function
\begin{align}
    \chi_{\rho}(\bm{r})=\Tr\qty[\hat D(\bm{r})\rho].
\end{align}
The characteristic function $\chi_{\rho}: \mathds{R}^{2m}\rightarrow \mathds{C}$ is the symplectic Fourier transform of the Wigner function. The characteristic function fulfills the requirement for a state $\Tr\rho$ that $\chi_{\rho}(0)=1$ and for $\rho^\dagger =\rho$ that $\chi_{\rho}(-\bm{r})= \chi_{\rho}(\bm{r})^*$.
The characteristic function of a Gaussian state with mean $\bm{\Bar{r}}$ and covariance matrix $\sigma$ is given by
\begin{align}
    \chi_{\rho_G}(\bm{r})=\e^{\frac{1}{4}\bm{r}^T\Omega^T \sigma \Omega \bm{r}-\ii \bm{\Bar{r}}\Omega \bm{r}}.
\end{align}
Using the characteristic function one can represent the action of a general Gaussian channel $\Phi_G$--- the maps that map Gaussian states to Gaussian states---
\begin{align}
    \chi_{\Phi(\rho)}(\bm{r})=\e^{-\frac{1}{4}\bm{r}^T \Omega^T Y \Omega \bm{r}-\ii \bm{\Bar{r}}\Omega \bm{r}}\chi_{\rho} \qty(\Omega^T X^T \Omega \bm{r})
\end{align}
with the condition.
\begin{align}
    Y+i\Omega\geq i X\Omega X^T.
\end{align}

\section{Monotonicity}
\label{ap:monotonicity}
In this section, we present the fundamental properties of our construction. These properties can then be used to establish monotonicity, once a measure is chosen. Specifically, we demonstrate which operations are equivalent to operations that act locally after the beam splitter.

\subsection{$n$ mode  to $n$ mode Gaussian channels}
In this section we show monotonicity of the measures using our construction under $n$ mode to $n$ mode Gaussian channels. In later sections we will show monotonicity under partial trace and Gaussian state preparation, which then directly implies monotonicity under Gaussian channels.
We first consider a $n$ mode to $n$ mode Gaussian channel without displacements
\begin{align}
    \chi_{\Phi_G(\rho)}(\bm{r})=\e^{-\frac{1}{4}\bm{r}^T \bar{Y} \bm{r}}\chi_{\rho} \qty(\bar{X}^T\bm{r})
\end{align}
with the short-hand notation $\Bar{Y}=\Omega^T Y \Omega $, $\bar{X}^T=\Omega^T X^T\Omega$. We deal with displacements separately later in the section.

If we apply the Gaussian channel to both output systems of the beam splitter, we have 
\begin{align}
    \chi_{\Phi(\rho)} \qty(\frac{\bm{r_A}+\bm{r_B}}{\sqrt{2}})     \chi_{\Phi(\rho)} \qty(\frac{\bm{r_B}-\bm{r_A}}{\sqrt{2}})&= \e^{-\frac{1}{4} \bm{r_A}^T\bar{Y}\bm{r_A}} \e^{-\frac{1}{4} \bm{r_B}^T\bar{Y}\bm{r_B}} \chi_{\rho} \left( \bar{X}^T\frac{\bm{r_A}+\bm{r_B}}{\sqrt{2}} \right)  \chi_{\rho} \left( \bar{X}^T\frac{\bm{r_B}-\bm{r_A}}{\sqrt{2}} \right).
\end{align}
If we now instead apply the channel before the beam splitter we have
\begin{align}
    \chi_{\Phi(\rho)}(\bm{r_A}) \chi_{\Phi(\rho)}(\bm{r_B})= \e^{-\frac{1}{4} \bm{r_A}^T\bar{Y}\bm{r_A}} \e^{-\frac{1}{4} \bm{r_B}^T\bar{Y}\bm{r_B}} \chi_{\rho}(\bar{X}^T\bm{r_A})  \chi_{\rho}(\bar{X}^T\bm{r_B}).
\end{align}
If we now apply the beam splitter we get
\begin{align}
    \chi_{\hat U_{BS}\Phi(\rho)\otimes \Phi(\rho) \hat U^\dagger_{BS} }(\bm{r_A},\bm{r_B})=\e^{-\frac{1}{4} \qty(\bm{r_A},\bm{r_B})S_{BS}^T \bar{Y}\oplus \bar{Y} S_{BS}\qty(\bm{r_A},\bm{r_B})^T } \chi_{\rho \otimes \rho }\qty(S_{BS} \bar{X}^T\oplus \bar{X}^T (\bm{r_A},\bm{r_B})^T).
\end{align}
It holds that 
\begin{align}
    \begin{pmatrix}
        \mathds{1}_m&\mathds{1}_m\\
        -\mathds{1}_m&\mathds{1}_m
    \end{pmatrix} \begin{pmatrix}
        \bar{X}^T&0\\
       0&\bar{X}^T
    \end{pmatrix} \begin{pmatrix}
        \bm{r_A}\\
        \bm{r_B}
    \end{pmatrix}= \begin{pmatrix}
         \bar{X}^T (\bm{r_A}+\bm{r_B})\\
        \bar{X}^T (\bm{r_B}-\bm{r_A})
    \end{pmatrix}
\end{align}

and for the exponential term
\begin{align}
    \qty(\bm{r_A},\bm{r_B})S_{BS}^T \bar{Y}\oplus \bar{Y} S_{BS}\qty(\bm{r_A},\bm{r_B})^T= \qty(\bm{r_A},\bm{r_B}) \bar{Y}\oplus \bar{Y}  \qty(\bm{r_A},\bm{r_B})^T.
\end{align}

We can thus multiply out the characteristic function and get
\begin{align}
    \chi_{\hat U_{BS}\Phi(\rho)\otimes \Phi(\rho) \hat U^\dagger_{BS} }(\bm{r_A},\bm{r_B})= e^{-\frac{1}{4} \bm{r_A}^T\bar{Y}\bm{r_A}} \e^{-\frac{1}{4} \bm{r_B}^T\bar{Y}\bm{r_B}} \chi_{\rho} \left( \bar{X}^T\frac{\bm{r_A}+\bm{r_B}}{2} \right ) \chi_{\rho} \left( \bar{X}^T\frac{\bm{r_B}-\bm{r_A}}{2} \right),
\end{align}
which is exactly the same as applying the channel after the beam splitter. We can thus commute the Gaussian channel through the beam splitter.
We are left to show how we can commute displacement operators through the beam splitter.
It holds that
\begin{align}
    \hat{D}(\bm{\bar{r}})\otimes \hat{D}(\bm{\bar{r}}) \hat U_{BS} = \hat U_{BS}  \hat{D}(\sqrt{2}\bm{\bar{r}})\otimes \hat{D}(0). 
\end{align}

Thus we have shown that Gaussian $n$ mode to $n$ mode channels can be moved through the beam splitter to act locally on both outputs. This cannot increase the total correlation.
Note that if $Y=0$, the matrix $X$ is a symplectic matrix and the channel describes Gaussian unitary evolution.

\subsection{Tracing out subsystem}
In this section we will show monotonicity under partial trace.
We start by considering as the input a state out of which a sub system $C$ is traced out.
The input state is then given as $\sigma=\Tr_C\qty[\rho]$, while the output  of the beam splitter is 
\begin{align}
    \chi_{\hat U_{BS}\qty(\sigma\otimes \sigma)\hat U_{BS}^\dagger}(\bm{r_A},\bm{r_B})&= \chi_{\Tr_C\qty[\rho]}\qty(\frac{\bm{r_A}+\bm{r_B}}{\sqrt{2}}) \chi_{\Tr_C\qty[\rho]}\qty(\frac{\bm{r_B}-\bm{r_A}}{\sqrt{2}})\\
    &= \chi_{\rho}\qty(\frac{\bm{r_A}+\bm{r_B}}{\sqrt{2}},\bm{r_C}=0)  \chi_{\rho}\qty(\frac{\bm{r_B}-\bm{r_A}}{\sqrt{2}},\bm{r_C}=0).
\end{align}
This means that the output state with the input $(\Tr_C[\rho])^{\otimes 2}$ coincides with the state obtained by tracing out $C$ systems of the output with input $\rho^{\otimes 2}$, whose characteristic function is
\begin{align}
     \chi_{\Tr_C \Tr_C'\hat U_{BS}\qty(\rho\otimes \rho)\hat U_{BS}^\dagger}(\bm{r_A}\oplus \bm{r_C},\bm{r_B}\oplus \bm{r_{C'}})&=\chi_{U_{BS}\qty(\rho\otimes \rho)\hat U_{BS}^\dagger}  (\bm{r_A}\oplus \bm{r_C}=0,\bm{r_B}\oplus \bm{r_{C'}=0})\\
     &=\chi_{\rho}\qty(\frac{\bm{r_A}+\bm{r_B},\bm{r_C=0}+\bm{r_{C'}=0}}{\sqrt{2}}  )
     \chi_{\rho}\qty(\frac{\bm{r_B}-\bm{r_A},\bm{r_{C'}=0}-\bm{r_{C}=0}}{\sqrt{2}}  )\\
     &=\chi_{\rho}\qty(\frac{\bm{r_A}+\bm{r_B}}{\sqrt{2}\bm{r_C}=0}  )\chi_{\rho}\qty(\frac{\bm{r_B}-\bm{r_A}}{\sqrt{2}},\bm{r_C}=0).
\end{align}
Thus tracing out before the beam splitter is equivalent to tracing out the outputs locally, the correlations cannot increase.

\subsection{Adding Gaussian subsystem}
We note that the output of tensor product states in our setup is a tensor product of the separate subsystems. As Gaussian states are product states after the beam splitters, we can add Gaussian subsystems without increasing the correlation of the output state.

\subsection{Multiplicative output}
In this section we show that if the input states are product states then the output will be multiplicative along the same axis. 
 We take two states of the form $\rho\otimes \sigma$ as the input of the beam splitter.
 The characteristic function before the beam splitter is
 \begin{align}
     \chi_{(\rho\otimes \sigma)\otimes (\rho\otimes \sigma)}(\bm{r_A},\bm{r_B})=\chi_{\rho\otimes \sigma}(\bm{r_A}) \chi_{\rho\otimes \sigma}(\bm{r_B})
 \end{align}
 and after the beam splitter
 \begin{align}
     \chi_{\rho\otimes \sigma}\qty(\frac{\bm{r_A}+\bm{r_B}}{\sqrt{2}}) \chi_{\rho\otimes \sigma}\qty(\frac{\bm{r_B}-\bm{r_A}}{\sqrt{2}})= \chi_{\rho}\qty(\frac{\bm{r_A^1}+\bm{r_B^1}}{\sqrt{2}})\chi_{\sigma}\qty(\frac{\bm{r_A^2}+\bm{r_B^2}}{\sqrt{2}}) \chi_{\rho}\qty(\frac{\bm{r_B^1}-\bm{r_A^1}}{\sqrt{2}}) \chi_{\sigma}\qty(\frac{\bm{r_B^2}-\bm{r_A^2}}{\sqrt{2}}).
 \end{align}
 At the same time, we have
\begin{align}
    \chi_{(\hat U_{BS}\rho \otimes \rho \hat U_{BS}^\dagger)\otimes (\hat U_{BS}\sigma \otimes \sigma \hat U_{BS}^\dagger)}(\bm{r_A^1},\bm{r_B^1},\bm{r_A^2},\bm{r_B^2})&= \chi_{\rho \otimes \rho}\qty(\frac{\bm{r_A^1}+\bm{r_B^1}}{\sqrt{2}}, \frac{\bm{r_B^1}-\bm{r_A^1}}{\sqrt{2}})\chi_{\sigma \otimes \sigma}\qty(\frac{\bm{r_A^2}+\bm{r_B^2}}{\sqrt{2}}, \frac{\bm{r_B^2}-\bm{r_A^2}}{\sqrt{2}}),
\end{align}
 which is equivalent to the case where we used the states $\rho\otimes \sigma$ as the input.
 The output states are thus the same as the output of two separate beam splitters.

This can be a very useful property depending on the measure of correlation employed.
Some measures such as have the multiplicative property.
Other measures do not have this property, but the fact the states factorizes can be used to derive useful bounds.

\section{Correlation without entanglement}
\label{ap:corrnotent}
In this work, we measure the amount of non-Gaussianity in two copies of the state by quantifying the correlations of the state after a $50{:}50$ beam splitter.
For pure states, the correlations are equivalent to the entanglement in the output state. 
This is not true for mixed states. If one wants a faithful measure one has to consider all types of correlations of the output state.
In the following we will provide a non-Gaussian state whose state psot beam splitter is a seperable state.
More specifically, we consider states of the form
\begin{align}
    \rho=\sum_i p_i \hat D(\bm{r_i}) \rho_G \hat D(\bm{r_i})^\dagger,
\end{align}
where the Gaussian state $\rho_G$ have zero mean.
Then it holds that
\begin{align}
   \hat  U_{BS} \rho \otimes \rho \hat U_{BS}^\dagger &= \sum_{i,j} p_i p_j  \hat U_{BS} \qty[\hat  D(\bm{r_i}) \rho_G \hat D(\bm{r_i})^\dagger \otimes \hat D(\bm{r_j}) \rho_G \hat D(\bm{r_j})^\dagger ]\hat U_{BS}^\dagger\\
      =&\sum_{i,j} p_i p_j   \hat U_{BS} \qty(\hat  D(\bm{r_i}) \otimes \hat  D(\bm{r_j}))\hat U_{BS}^\dagger U_{BS}\rho_G \otimes \rho_G \hat U_{BS}^\dagger \hat U_{BS}\qty(\hat  D(\bm{r_i}) \otimes \hat  D(\bm{r_j}))^\dagger \hat U_{BS}^\dagger\\
    &= \sum_{i,j} p_i p_j   \hat U_{BS} \qty(\hat  D(\bm{r_i}) \otimes \hat  D(\bm{r_j}))\hat U_{BS}^\dagger \rho_G \otimes \rho_G  \hat U_{BS}\qty(\hat  D(\bm{r_i}) \otimes \hat  D(\bm{r_j}))^\dagger \hat U_{BS}^\dagger\\
    &=\sum_{i,j}p_ip_j  \qty(\hat D\qty(\frac{\bm{r_i}-\bm{r_j}}{\sqrt{2}}) \otimes \hat D\qty(\frac{\bm{r_i}+\bm{r_j}}{\sqrt{2}})) \rho_G \otimes \rho_G \qty(\hat D\qty(\frac{\bm{r_i}-\bm{r_j}}{\sqrt{2}})^\dagger \otimes \hat D\qty(\frac{\bm{r_i}+\bm{r_j}}{\sqrt{2}})^\dagger). 
\end{align}
In the second line we used that $\hat U_{BS} \rho_G\otimes \rho_G \hat U_{BS}^\dagger = \rho_G\otimes \rho_G$ if $\rho_G$ has zero mean, i.e. $\mu$.
Indeed we obtain a separable state.

\section{Examples}
\label{ap:examples}
In this section we provide examples for the amount of non-Gaussianity measured by the correlation after the beam splitter.
As we are considering pure state we will use the entanglement monotones the von Neumann entropy and the 2-Rényi entropy~\cite{PhysRevA.93.022324}.

The Schmidt decomposition of a pure state $\psi$ is given by
\begin{align}
    \ket{\psi}=\sum_i \sqrt{\lambda_i} \ket{i}_A\otimes \ket{i}_B.
\end{align}
The entanglement Rényi-$\alpha$ entropy~\cite{PhysRevA.93.022324} is then 
\begin{align}
    E_{\alpha}(\psi)=\frac{1}{1-\alpha}\log_2\qty(\sum_i \lambda_i^\alpha)
\end{align}
 with the special case $\alpha=1$
 \begin{align}
     E_{1}(\psi)=-\sum_i \lambda_i \log_2 \lambda_i
 \end{align}
and
\begin{align}
    E_2(\psi)=-\log_2\qty(\sum_i\lambda_i^2).
\end{align}

If we have two copies of pure state $\psi$ in the input of the beam splitter then the output is $\chi_\psi(\frac{\bm{r_2}+\bm{r_1}}{\sqrt{2}})\chi_\psi(\frac{\bm{r_2}-\bm{r_1}}{\sqrt{2}})$. The reduced state is then $\chi_{\rho}= \chi_\psi(\frac{\bm{r_1}}{\sqrt{2}}) \chi_\psi(\frac{-\bm{r_1}}{\sqrt{2}})$ and thus
\begin{align}
    \Tr\qty[\rho^2]=\frac{1}{(2\pi)^n} \int \dd\bm{r} \chi_\rho(\bm{r}) \chi_\rho^*(\bm{r})=\frac{1}{(2\pi)^n} \int \dd\bm{r} \abs{\chi_{\psi}\qty(\frac{\bm{r}}{\sqrt{2}})}^4.
\end{align}
In Figure~\ref{fig:E2} we show the entanglement $2$-Rényi entropy for Fock states, $0N$ states and cubic phase states.
The $0N$ state we are considering is defined as a equal superposition between the vacuum and a Fock state $\ket{N}$
\begin{align}
    \ket{0N}=\frac{1}{\sqrt{2}}\qty(\ket{0}+\ket{N}).
\end{align}
In Appendix~\ref{ap:lower_bound}, we discuss additional properties of the cubic phase state; for convenience, we repeat its definition here:
\begin{align}
\ket{\gamma, r} = e^{i \gamma \hat{q}^3} \hat{S}(r) \ket{0}.
\end{align}

\subsection{Fock states}
Here, we derive analytical expressions quantifying the non-Gaussianity of Fock states.
The output state for two identical Fock states after the beam splitter is~\cite{PhysRevA.65.032323}
\begin{align}
   \ket{\Psi}= \hat U_{BS}\ket{n,n} &= \sum_{m,k=0}^n (-1)^{n-k} \binom{n}{k} \binom{n}{2m-k} \frac{\sqrt{2m! (2n-2m)!}}{n!} \ket{2m,2n-2m}\\
    &= \sum_{m=0}^n c_m^n \ket{2m,2n-2m}
\end{align}
with 
\begin{align}
    c_m^n=\frac{1}{2^n} \sum_{k=0}^n (-1)^{n-k} \binom{n}{k}\binom{n}{2m-k}\frac{2m! (2n-2m)!}{n!}.
\end{align}
Thus the non-Gaussianity measured using the  entanglement entropy is
 \begin{align}
     N_{E_1}(\ket{n})=-\sum_{m=0}^n \abs{c_m^n}^2 \log_2 \abs{c_m^n}^2,
 \end{align}
 and the 2-Rényi entropy
 \begin{align}
     N_{E_2}(\ket{n})=-\log_2\qty(\sum_{m=0}^n \abs{c_m^n}^4 ).
 \end{align}
 
\subsection{Cat states}
In this section we are providing analytical expression for squeezed cat states
of the form $\ket{\psi}=\sum_{i=0}^1 c_i \hat S(s)\ket{\alpha_i}$, where $\hat S(s)$ is the squeezing operator and a coherent state $\ket{\alpha_i}$ with real $\alpha_i$. See Appendix~\ref{ap:background} for the precise definitions.
The output after the beam splitter is given by
\begin{align}
    \ket{\Psi}=\hat U_{BS}\ket{\psi}\otimes \ket{\psi}&=\sum_{i,j=0}^{1} c_i c_j  \hat U_{BS} \hat S(s)\otimes\hat S(s)\ket{\alpha_i, \alpha_j}\\
    &=\sum_{i,j=0}^{1} c_i c_j  \hat U_{BS} \hat S(s)\otimes\hat S(s) \hat U_{BS}^\dagger   \hat U_{BS}  \hat{D}(\alpha_i)\otimes  \hat{D}(\alpha_i) \hat U_{BS} ^\dagger \hat U_{BS} \ket{0,0}\\
    &=\sum_{i,j=0}^{1} c_i c_j \hat S(s)\otimes \hat S(s)\ket{\frac{1}{2}(\alpha_j+\alpha_i), \frac{1}{\sqrt{2}}(\alpha_j-\alpha_i)}\\
    &=\sum_{\bm{k}} \Bar{c}_{\bm{k}} \hat S(s)\otimes \hat S(s) \ket{\Bar{\alpha}_{\bm{k}},\Bar{\alpha}_{\bm{k}}'},
\end{align}
with $\bm{k}=\qty(i,j)$, $\Bar{c}_{\bm{k}}=c_i c_j$, $\Bar{\alpha}_{\bm{k}}=\frac{1}{\sqrt{2}}(\alpha_j+\alpha_i)$, and $\Bar{\alpha}_{\bm{k}}'=\frac{1}{2}(\alpha_j-\alpha_i)$, where in the third line we used that the squeezing can be commuted through the beam splitter.

As long as $\abs{\Bar{\alpha}_{\bm{k}}-\Bar{\alpha}_{\bm{k}'}}\gg 1$ or $\abs{\Bar{\alpha}_{\bm{k}}'-\Bar{\alpha}_{\bm{k}'}'}\gg 1$ all $\ket{\Bar{\alpha}_{\bm{k}},\Bar{\alpha}_{\bm{k}}'}$ are approximately orthogonal.
This holds for standard cat states with large enough $\alpha$.
The non-Gaussianity measured through the entanglement entropy and 2-Rényi entropy can then be approximated as 
 \begin{align}
    N_{E_1}(\ket{\psi})\sim -\sum_{\bm{k}}  \abs{\Bar{c}_{\bm{k}}}^2 \log_2 \abs{\Bar{c}_{\bm{k}}}^2,
 \end{align}
 \begin{align}
     N_{E_2}(\ket{\psi})\sim -\log_2\qty(\sum_{m=0}^n  \abs{\Bar{c}_{\bm{k}}}^2 ).
 \end{align}

\section{The correlation cannot be detected by covariance matrix}
\label{ap:not_in_cov}
In this section, we show that the correlation generated by beam splitter cannot be detected solely through the covariance matrix. While every quantum state admits an associated covariance matrix, only Gaussian states are fully characterized by their first and second moments—that is, by their mean vector and covariance matrix.

Recall that the symplectic transformation corresponding to the beam splitter unitary is given by:
\begin{align}
        S_{BS}=\frac{1}{\sqrt{2}}\begin{pmatrix}
        \mathds{1}_m&\mathds{1}_m\\
        -\mathds{1}_m&\mathds{1}_m
    \end{pmatrix}.
\end{align}

The covariance matrix of the two copies of the input state of the beam splitter is
\begin{align}
    \sigma =\begin{pmatrix}
        \sigma_{\rho}&0\\
        0 &  \sigma_{\rho}
    \end{pmatrix}
\end{align}
for a general---not necessarily Gaussian---state $\rho$. 
$\sigma_{\rho}= \Tr\qty[   \{ \hat{\bm{r}}-\mu,   \qty(\hat{\bm{r}}-\mu)^T \} \rho]$ is the covariance matrix of the state $\rho$, where $\mu$ is the mean of the state $\rho$ and $\{,\}$ the anti-commutator. 

We then track how the covariance matrix is transformed through the beam splitting interaction.
The covariance matrix of the output is 
\begin{align}
    S_{BS}\sigma S_{BS}^T = \begin{pmatrix}
        \sigma_{\rho}&0\\
        0 &  \sigma_{\rho}
    \end{pmatrix} =\sigma.
\end{align}
We can see that the covariance matrix before and after the beam splitter is identical. Thus, we cannot measure or detect the correlation by only considering the covariance matrix.

\section{Measuring mutual information}
\label{ap:meas_mutual}
Experimentally accessing the Rényi-$\alpha$ mutual information remains an open problem. Nevertheless, we provide a few comments on possible avenues for its measurement. Although its definition involves a minimization, the bounds established in~\cite{10.1063/1.5143862} can be used to obtain estimates of the resource content.
Theorem 1 in Ref.~\cite{10.1063/1.5143862} states that for $\alpha,\beta>0$ and $\gamma\geq\frac{1}{2}$  satisfying $\frac{\alpha}{\alpha-1}=\frac{\beta}{\beta-1}+\frac{\gamma}{\gamma-1}$ that for $(\alpha-1)(\beta-1)(\gamma-1)>0$ 
\begin{align}
    I_{\alpha}^{\uparrow}(A;B)_\rho&\geq H_\beta(B)_\rho-H_{\alpha}^{\downarrow}(B\lvert A)_\rho\\
      I_{\alpha}^{\downarrow}(A;B)_{\rho}&\geq H_\beta(B)_{\rho}-H_{\alpha}^{\uparrow}(B\lvert A)_{\rho}
\end{align}
and for $(\alpha-1)(\beta-1)(\gamma-1)<0$ 
\begin{align}
    I_{\alpha}^{\uparrow}(A;B)_{\rho}&\leq H_\beta(B)_{\rho}-H_{\alpha}^{\downarrow}(B\lvert A)_{\rho}\\
      I_{\alpha}^{\downarrow}(A;B)_{\rho}&\leq H_\beta(B)_{\rho}-H_{\alpha}^{\uparrow}(B\lvert A_{\rho}).
\end{align}
We can use the expression provided in~\cite{Tomamichel_2016} to evaluate the bounds:
\begin{align}
    H_{\alpha}^{\uparrow}(A\lvert B)=\frac{\alpha}{1-\alpha}\log_2\Tr\qty[\qty(\Tr_A(\rho^\alpha_{AB}))^\frac{1}{\alpha}]
\end{align}
for all $\alpha\in (0,1)\cup (1,\infty)$.
The conditional entropy $H^{\downarrow}$ has a convenient form for $\alpha=2$
\begin{align}
    H^{\downarrow}_2(A\lvert B)_{\rho}=-\log_2\qty[  \rho_{AB}  \qty(\mathds{1}_A\otimes \rho_B^{-\frac{1}{2}})\rho_{AB}\qty(\mathds{1}_A\otimes \rho_B^{-\frac{1}{2}})].
\end{align}
It could be possible to estimate these entropies using the quantum singular value transform~\cite{10.1145/3313276.3316366}.
We leave a specific protocol for future work.

\section{Classical shadow}
\label{ap:shadows}
Here, we summarize the use of classical shadows in our framework.
Classical shadows are a powerful framework for quantum state estimation with a wide range of applications~\cite{Huang_2022}. Beyond state tomography, classical shadows can be used to estimate quantities such as the von Neumann entropy and the purity of subsystems~\cite{Huang_2022}.

Unlike the techniques discussed earlier in this supplemental material—which require simultaneous access to multiple copies of the quantum state—classical shadows only require sequential access to individual copies. This makes them particularly appealing for experimental implementations.
This section follows the approach introduced in Ref.~\cite{Becker_2024}.

The shadows are obtained by using homodyne detection along a random axis.
Assume we want to obtain classical shadows for an $m-$mode quantum state $\rho$.
We sample Haar random $m$ matrices $\text{Sp}(2) \cap \text{SO}(2)\cong \text{U}(1)$ $S_i$.
To this end, we sample random angles $\theta_1,\dots,\theta_m$ to get 
\begin{align}
    U_{S_j} =R(\theta_j)=\begin{pmatrix}
        \cos(\theta_j)& -\sin(\theta_j)\\
        \sin(\theta_j)&  \cos(\theta_j)
    \end{pmatrix}
\end{align}
Then we use a copy of the state $\rho$ and apply
the rotation $\hat U_{S}\rho \hat U_S^\dagger$ with $\hat U_S=\hat U_{S_1}\otimes ...\otimes \hat U_{S_m}$ and thus one random rotation or one sampled angle per mode.
Afterwards we perform a homodyne measurement along the position axis and obtain $\bm{x}=\qty(\bm{x_1},...,\bm{x_m})^T\in \mathds{R}^{2m}$.
The effect of averaging over all random matrices $S$ and all measurement outcomes $\bm{x}$ 
and undoing the transformation of $S$ in the post-measurement state $\Tilde{\rho}_x$, acts like the quantum channel
\begin{align}
    \mathcal{E(\rho)}=\mathds{E}_{S,x}\qty[\hat U_S^\dagger \Tilde{\rho}_x \hat U_S].
\end{align}
For homodyne detection, $\mathcal{E}$ is a simple linear bosonic channel, whose action can be represented as a random displacement channel. The channel is injective and invertible on its range.

We then define the classical shadow as 
\begin{align}
    \bar{\rho}_{S,x}=\mathcal{E}^{-1}\qty(\hat U_S^\dagger \Tilde{\rho}_x \hat U_S).
\end{align}
$\bar{\rho}_{S,x}$ is an unbiased estimation of $\rho$. 
The information of $\rho$ is contained in the probability distribution of obtaining measurement outcome $x$.

We only need to perform the initial rotation and the measurement of the quantum state using a quantum device. 
The rest of the operation, such as the counter-rotation and the inverse of the channel, can be done in an efficient postprocessing, as everything involved is Gaussian.

In the case of homodyne detection, the post-measurement ``state'' of subsystem $A$ including the inverse of the channel is then given by
\begin{align}
    \chi_{\bar{\rho}^{(i)}}\qty(\bm{r_A})=\prod_{j\in A} \sqrt{2}\pi \norm{\bm{r_j}}\delta\qty((S_j\bm{r_j})_2)\e^{-\ii \bm{r_j}^T\Omega S_j \bm{x_j}}
\end{align}
for any $\bm{r_A}=\{\bm{r_j}\}_{j\in A}\in \mathds{R}^{2\abs{A}}$ and $(S_j\bm{r_j})_2=\sum_k (S_j)_{2k}(\bm{r_j})_k$. This characteristic function is not associated to a quantum state. 
We can however use it to construct a random matrix of size $(M+1)^\abs{A}$ by
\begin{align}
    \bar{\rho}_A^{(i)}(M)=\sum_{\bm{n_1},\bm{n_2}\in\{0,...,M   \}^A}\langle \chi_{\ket{\bm{n_1}}\bra{\bm{n_2}}},\chi_{\hat{\rho}^{(i)}}\rangle\ket{\bm{n_1}}\bra{\bm{n_2}}
\end{align}
where $\chi_{\ket{n_1}\bra{n_2}}$ are the matrix elements of the displacement operator in Fock basis.

By repeating this $N$ times and compute the average we obtain the classical shadow
\begin{align}
    \sigma^{(N)}(M)=\frac{1}{N}\sum_{i=1}^N \bar{\rho}^{(i)}(M).
\end{align}

\subsection{Non-linear functionals}

Our main interest is to use classical shadows to obtain non-linear functionals of the state $\rho$.
The examples we are interested are purity of a subsystem $P(\rho_A)=\Tr\qty[\rho_A^2]$ and the von Neumann entropy of a subsystems $S(\rho_A)=-\Tr\qty[\rho_A\log\rho_A]$. We consider only the reduced state $\rho_A$ where $A$ denotes the set of modes containing $\abs{A}=r$ modes with $r<m$. We assume that the state $\rho$ has local finite energy $\Tr\qty[(\mathds{1}+n^j)\rho_j]=E_j<\infty$ on mode $j$, which is true for any experiment.

\paragraph{von Neumann entropy:} We start with the von Neumann entropy. The derivation can be found in~\cite{Becker_2024}.
Given the shadow $\sigma_A^{(N)}(M)$ obtained from the scheme described in the previous section and a constant $d_p\in \mathds{N}$, we use the following matrix functional
\begin{align}
    H(d_p)(\sigma_A^{(N)}(M))=-\Tr\qty[\sigma_A^{(N)}(M)-P_M]+\sum_{k=2}^{d_p}\frac{\Tr\qty[(P_M-\sigma_A^{(N)}(M))]}{k(k-1)}
\end{align}
where $P_M$ projects into the subspace of dimensions $M+1$ that contains up to Fock number $M$ in order to approximate the entropy.
$M$ thus describes the highest Fock number one takes into account.
It holds then that for any $\epsilon>0$ and $M\in\mathds{N}$ that
\begin{align}
    {\rm Prob}\qty(\exists A, \abs{A}\leq r:\abs{S(\rho_A)-H\qty( \left\lceil\frac{3(M+1)^r}{\epsilon}  \right\rceil )(\sigma_A^{(N)}(M))}\geq \epsilon)\leq2\qty[m(M+1)]^r \exp\qty(-\frac{3N\epsilon^{'2}(M+1)^{-2r}}{6\Sigma_r^{(0)}(M)^2+2( \Sigma_r^{(0)}(M)+1)\epsilon'})
\end{align}
with $\epsilon'=\frac{\epsilon^2}{14(M+1)^r\e} 2^{-4(M+1)^r/\epsilon}$,
\begin{align}
    \Sigma_r^{(0)}=\norm{\qty( \prod_{j=1}^r \sqrt{\frac{n_2(j)!}{n_1(j)!}}  \int \dd y\abs{\sqrt{\pi}y}^{1+M_j-m_j}\e^{-\frac{\abs{y}^2}{8\pi}} \abs{L^{M_j-m_j}_{M_j}\qty(\pi \abs{y}^2)}  )_{\bm{n_1},\bm{n_2}}}_\infty
\end{align}
where $\bm{n_1},\bm{n_2}\in\{0,..,M\}^r$, 
$M_j=\max\{n_1(j),n_2(j)  \}$, $m_j=\max\{n_1(j),n_2(j)  \}$ and $L^{(k)}_j$ being Laguerre polynomials.

\paragraph{Purity:} The other non-linear functional we are interests in is the purity. 
We can use Theorem 3~\cite{Becker_2024} to upper bound the number of samples. 
If the energy of the state is bounded like $E_r=\max_{\abs{A}\leq r} \Tr\qty[\rho_A H_r]\leq \infty$, the number of samples is
\begin{align}
    N\geq \frac{(M+1)^{2r}}{3\epsilon^2}\qty(24\Sigma^{(0)}(M)^2+4\qty[\Sigma^{(0)}(M)+E_r]\epsilon)\log\qty(\frac{2L(M+1)^r}{\delta})
\end{align}
and \begin{align}
    M=\left\lceil \qty(\frac{4E_r}{\epsilon})^2 \right\rceil
\end{align}
then it holds for $L$ observables $O_j$ with $\norm{O_j}_\infty\leq 1$ on regions $A_j$ of size at most $r$  that
\begin{align}
    \max_j\abs{\Tr\qty[O_j\qty(\sigma^{(N)}_{A_j}-\rho_{A_j})]}\leq \epsilon
\end{align}
with probability at least $1-\delta$. 

A quantum state $\rho_A$ is an example of such an observable as it holds that $\norm{\rho_A}_\infty\leq 1$ and then $L=1$.

\section{Sampling cost lower bound for estimating Wigner negativity}
\label{ap:lower_bound}
We are interested in comparing the experimental difficulty of estimating non-Gaussianity using our protocol with that of estimating the Wigner negativity. To this end, we compute a lower bound on the number of samples required for Wigner negativity estimation.
The Wigner negativity of the state $\rho$ is defined by 
\begin{align}
    \mathcal{W}(\rho)=\int \dd\bm{r}\abs{W_{\rho}(\bm{r})}.
\end{align}
The standard approach to determining the Wigner negativity of a quantum state in an experiment involves full state or Wigner tomography, followed by reconstruction of the Wigner function. However, such complete reconstructions may be unnecessary when the goal is to estimate the Wigner negativity itself, which contains significantly less information than the full state.

\subsection{State discrimination}
We prove the lower bound on the number of samples required to obtain the Wigner negativity of a quantum state by connecting the task of estimating the Wigner negativity to a state discrimination task.
The optimal probability to succeed in discriminating between the states $\rho$ and $\sigma$ using $n$ copies is given by
\begin{align}
    \frac{1}{2}\qty( \frac{1}{2}\norm{\rho^{\otimes n}-\sigma^{\otimes n}}_1+1)=p_{\text{succ},n}.
\end{align}
Suppose we can estimate $\mathcal{W}(\rho)$ $\forall \rho$ with additive error $\epsilon$ and failure probability $1-\delta$ with samples $N(\epsilon,\delta)$.
Suppose 
\begin{align}
    \abs{\mathcal{W}(\rho)-\mathcal{W}(\sigma)}\geq 2 \epsilon
    \label{eq:Wigner negativity difference}
\end{align}
then we can discriminate $\rho$ and $\sigma$ with probability $1-\delta$ and $N(\epsilon,\delta)$ samples. This implies
\begin{align}
    \frac{1}{2}\qty(\frac{1}{2}\norm{\rho^{\otimes N(\epsilon,\delta)} -\sigma^{\otimes N(\epsilon,\delta)}}_1+1)\geq 1-\delta.
    \label{eq:lower bound trace distance}
\end{align}

In particular, if $\rho$ and $\sigma$ are pure states $\psi$ and $\phi$, we have $\norm{\psi^{\otimes n}-\phi^{\otimes n}}_1= 2\sqrt{1-F(\psi^{\otimes n},\phi^{\otimes n})}=2\sqrt{1-F(\psi,\phi)^n}$, which gives 
\begin{align}
    \frac{1}{2}\qty(\sqrt{1-F(\psi,\phi)^{N(\epsilon,\delta)}}+1)\geq 1-\delta.
\end{align}
Then
\bal
    1-F(\psi,\phi)^{N(\epsilon,\delta)}&\geq (1-2\delta)^2\\
    -F(\psi,\phi)^{N(\epsilon,\delta) }&\geq  (1-2\delta)^2-1\\
    F(\psi,\phi)^{N(\epsilon,\delta)}&\leq 1-(1-2\delta)^2\\
    N(\epsilon,\delta) \log(F(\psi,\phi)) &\leq 4\delta (1-\delta)\\,
\eal 
which provides a sampling lower bound for estimating the Wigner negativity as
\bal 
N(\epsilon,\delta) &\geq  \frac{\log(\frac{1}{4\delta (1-\delta)})}{\log(\frac{1}{F(\psi,\phi)})}.
\label{eq:sampling lower bound negativity}
\eal

In the following, we aim to put a general lower bound for the number of samples required to estimate the Wigner negativity of an arbitrary unknown state. 
We achieve this by putting a lower bound for $N(\epsilon,\delta)$ via \eqref{eq:sampling lower bound negativity} for a specific choice of $\psi$ and $\phi$ satisfying \eqref{eq:Wigner negativity difference}, which serves as a lower bound for the sampling cost required for estimating arbitrary unknown states.
To this end, we consider a class of states known as cubic phase states, which we discuss in the next subsection.

\subsection{Cubic phase state}
The general bound derived above can now be made explicit by considering concrete examples. In what follows, we focus on the case of cubic phase state.
This section provides a summary of the most important properties of the cubic phase state as well as a lower bound on their Wigner negativity.

The cubic phase state,
\begin{align}
\ket{\gamma, r} = e^{i \gamma \hat{q}^3} \hat{S}(r) \ket{0},
\end{align}
is a paradigmatic example of a highly non-Gaussian quantum state. It plays a central role in continuous-variable quantum computation, where it enables universal gate sets. In particular, the cubic phase state allows for the implementation of a non-Clifford $T$-gate on the Gottesman-Kitaev-Preskill (GKP) code space via gate teleportation~\cite{PhysRevA.64.012310}. More generally, it also features in the Braunstein–Lloyd scheme~\cite{PhysRevLett.82.1784}, where teleporting a (possibly noisy or Gaussian-diluted) cubic phase gate provides the necessary nonlinearity for universal quantum computation.

The fidelity between two cubic phase states with same $\gamma$ is given as~\cite{PhysRevA.97.062337}
\bal
    F(\ket{c,r},\ket{c,r'}) &= \abs{\bra{0}\hat S(r)^\dagger \e^{-i c \hat q^3} \e^{i c \hat q^3} \hat S(r')\ket{0}}^2\\
    &=\abs{\bra{0} \hat S(r)^\dagger \hat S(r')\ket{0}}^2\\
    &= \frac{2}{\sqrt{\det\qty(\sigma_r+\sigma_{r'})}}\\
    &= \frac{1}{\cosh(r-r')}.
    \label{eq:cubic phase state fidelity}
\eal
We see that the fidelity between cubic phase states with same $\gamma$ only depends on the squeezing difference. This implies we can increase the difference in Wigner negativity by increasing $c$ and $r$ while keeping the fidelity and thus $r-r'$ constant.

The Wigner function of the cubic phase state $\ket{\gamma,r}$ is given by~\cite{PhysRevA.97.062337,moore2025nonlinearphasegatesairy}
\begin{align}
    W(q,p)=\frac{\e^{-q^2/2\e^{-2r}}}{2\sqrt{\pi}} \frac{1}{\frac{1}{2}\sqrt{\frac{\e^{-2r}}{2}}(6\gamma)^{\frac{1}{3}}}\e^{\frac{2}{\sqrt{\frac{\e^{-2r}}{2}}^3  6\gamma}\qty(\sqrt{\frac{\e^{-2r}}{2}} (p+3\gamma q^2)  +\frac{1}{3\sqrt{\frac{\e^{-2r}}{2}}^3 6\gamma} )} \text{Ai}\qty(\frac{p+3\gamma q^2}{\frac{1}{2} (6\gamma)^{\frac{1}{3}}}+ \frac{1}{\sqrt{\frac{\e^{-2r}}{2}}^4 (6\gamma)^\frac{4}{3}}  ).
\end{align}

It has to be noted that there are many cubic phase states that will have the same Wigner negativity.
As is noted in~\cite{PhysRevA.98.052350} it holds that
\begin{align}
    \hat S(-r')\ket{\gamma, r+r'} = \ket{\gamma \e^{3r},r}.
\end{align}
The Wigner negativity of a state $\rho$ is defined by 
\begin{align}
    \mathcal{W}(\rho)=\int \dd\bm{r}\abs{W_{\rho}(\bm{r})}.
\end{align}
Since the Wigner negativity is invariant under Gaussian unitaries, Wigner negativity is constant on lines $x=\gamma \e^{3r}$ 
\begin{align}
    \mathcal{W}(\ket{\gamma ,r})=  \mathcal{W}(\ket{\gamma \e^{3r},0})=: \mathcal{W}(\e^{3r}\gamma).
\end{align}

\begin{figure}
    \centering
    \includegraphics[width=0.6\linewidth]{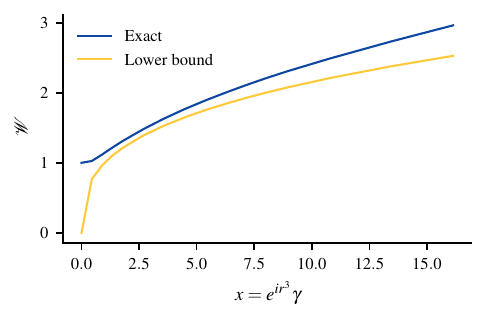}
    \caption{Plot showing how the Wigner negativity increases with higher $e^{3r}\gamma$. We see that for large $x$ that the Wigner negativity grows faster than the lower bounding function $f(x)=x^{\frac{1}{3}}$.}
    \label{fig:cubc_neg}
\end{figure}

In Figure~\ref{fig:cubc_neg} we show the Wigner negativity for increasing $x=\e^{3r}\gamma$. We see that $\mathcal{W}(\e^{3r}\gamma)$ can be lower bounded by the function $f(x)=x^{\frac{1}{3}}$. Importantly, for large $x$, we observe that the true Wigner negativity grows faster than $x^{\frac{1}{3}}$.
This is especially useful for values of $x$ that we cannot compute numerically any longer.
We thus want to emphasize that this heuristic lower bound is applicable for large $x$.  

As we noted above, there are many cubic phase states with the same Wigner negativity. We are interested in which of those cubic phase states with constant $x$ has the lowest photon number.
The mean photon number of a cubic phase state is given by~\cite{PhysRevA.97.062337}
\begin{align}
    E=\frac{1}{2}\qty(\cosh{2r}-1)+18 \gamma^2 \e^{4r}+\frac{1}{4}\qty(P+6\gamma\e^{2r})^2,
\end{align}
where we allowed for a displacement of $P$. Displacing the cubic phase state does not change the Wigner negativity and thus we choose $P= - 6\gamma \e^{2r}$.
We will write $x=\gamma \e^{3r}$ and thus
\begin{align}
    E=\frac{1}{2}\qty(\cosh{2r}-1)+18 x^2 \e^{-2r}.
\end{align}
We set $x$ constant and see which $r$ minimizes the mean photon number.
The first derivative is
\begin{align}
     \sinh\qty(2r)-36 x^2\e^{-2r}&=0\\
    \e^{4r}&=72x^2+1\\
    r&=\frac{1}{4}\log\qty(72x^2+1).
\end{align}
 The minimal mean photon number for a given value of $x$ is then
\begin{align}
    E(x)=\frac{1}{2}\qty(\frac{36x^2+1}{\sqrt{72x^2+1}}-1)+\frac{18x^2}{\sqrt{72x^2+1}}.
\end{align}
\subsubsection{Lower bound for Wigner negativity}
While an analytic expression for the Wigner negativity is not known, and numerical evaluation is limited for large $x$, asymptotic estimates provide meaningful insight.
In this section we provide a lower bound on the Wigner negativity in the limit of large $x$.
More specifically, we want to find a lower bound of
\begin{align}
    \norm{W_\psi}_1=\int \dd q \dd p \frac{\e^{-q^2/2}}{2\sqrt{\pi}} \frac{1}{\frac{1}{2}\sqrt{\frac{1}{2}}(6x)^{\frac{1}{3}}}
    \e^{\frac{2}{\sqrt{\frac{1}{2}}^3  6x}\qty(\sqrt{\frac{1}{2}} (p+3x q^2)  +\frac{1}{3\sqrt{\frac{1}{2}}^3 6x} )} 
        \abs{\text{Ai}\qty(\frac{p+3x q^2}{\frac{1}{2} (6x)^{\frac{1}{3}}}+ \frac{1}{\sqrt{\frac{1}{2}}^4 (6x)^\frac{4}{3}}  )}.
\end{align}

For $x$ large enough, the term $\frac{1}{\sqrt{\frac{1}{2}}^4 (6x)^\frac{4}{3}}$ does not contribute
in $\text{Ai}\qty(\frac{p+3x q^2}{\frac{1}{2} (6x)^{\frac{1}{3}}}+ \frac{1}{\sqrt{\frac{1}{2}}^4 (6x)^\frac{4}{3}}  )$.
Then
\begin{align}
    \norm{W_\psi}_1&\propto \frac{1}{x^{\frac{1}{3}}}\int_{-\infty}^\infty \dd q \dd p\e^{-\frac{3}{2}q^2} \e^{p/3x} \e^{4/27x}\abs{\text{Ai}\qty(\frac{p+3x q^2}{\frac{1}{2} (6x)^{\frac{1}{3}}})}\\
    &\propto \frac{1}{x^{\frac{1}{3}}} \e^{4/27x}\int_{-\infty}^\infty \dd q \dd p\e^{-\frac{1}{2}(q^2-\frac{2}{3}p)/x^{\frac{2}{3}}}  \abs{\text{Ai}\qty(\frac{p+q^2}{\frac{1}{2} (6)^{\frac{1}{3}}})}
\end{align}
The Airy function decays faster than exponential for positive argument. For a positive argument and large $x$, the corresponding integral is independent of $x$. 
We then bound the integral by first neglecting the integration over $q$, since $q$ only appears quadratic in the argument of the Airy function
\begin{align}
    &\frac{1}{x^{\frac{1}{3}}} \e^{4/27x}\int_{-\infty}^\infty \dd q \dd p\e^{-\frac{1}{2}(q^2-\frac{2}{3}p)/x^{\frac{2}{3}}}  \abs{\text{Ai}\qty(\frac{p+q^2}{\frac{1}{2} (6)^{\frac{1}{3}}})}\\
    &\geq \frac{1}{x^{\frac{1}{3}}} \int_{-\infty}^\infty \dd q \dd p\e^{-\frac{1}{2}(q^2-\frac{2}{3}p)/x^{\frac{2}{3}}}  \abs{\text{Ai}\qty(\frac{p+q^2}{\frac{1}{2} (6)^{\frac{1}{3}}})}\\
    &\geq \frac{1}{x^{\frac{1}{3}}} \int_{-\infty}^\infty \dd p\e^{\frac{1}{3}p/x^{\frac{2}{3}}}  \abs{\text{Ai}\qty(\frac{p}{\frac{1}{2} (6)^{\frac{1}{3}}})}\\
    &\geq  \frac{1}{x^{\frac{1}{3}}} \int_{0}^\infty \dd p\e^{-\frac{1}{3}p/x^{\frac{2}{3}}}  \abs{\text{Ai}\qty(-\frac{p}{\frac{1}{2} (6)^{\frac{1}{3}}})}\\
    &= x^{\frac{1}{3}} \int_{0}^\infty \dd p \e^{-p/3} \abs{\text{Ai}\qty(-\frac{x^{\frac{2}{3}}p}{\frac{1}{2} (6)^{\frac{1}{3}}})}.
\end{align}
For large $x$ we use the asymptotic for negative arguments $\text{Ai}(x)\simeq \frac{\sin\qty(\frac{2}{3}x^{\frac{3}{2}}+\frac{\pi}{4})}{\sqrt{\pi}x^{\frac{1}{4}}}$.
Since we are only interested in lower bounding the Wigner negativity asymptotically for large $x$, we neglect constant factors and obtain
\begin{align}
  x^{\frac{1}{3}} \int_{a}^\infty \dd p \e^{-p/3} \abs{ \frac{\sin\qty( x p^{\frac{3}{2}}+\frac{\pi}{4})}{ x^{\frac{1}{6}}  p^{\frac{1}{4}}} }.  
\end{align}
For large $x$ the term $\sin$ oscillates very fast compared to the exponential decay. We can then approximate it with a constant $c$.
In the end we obtain
\begin{align}
     x^{\frac{1}{6}} \int_{a}^\infty \dd p \e^{-p/3} \frac{1}{p^{\frac{1}{4}}} \sim x^{\frac{1}{6}}.
\end{align}
We can therefore see that we obtain a crude lower bound for large $x$ as $\norm{W_\psi}_1 (x)\geq x^{\frac{1}{6}}$.
In any case, we observe that the Wigner negativity for large $x$ follows a power law.
As mentioned before in the section, we see numerically that the function $x^{\frac{1}{3}}$ is a better bound.

\subsection{Sampling lower bound using the cubic phase state}
Having established a general sample lower bound for estimating Wigner negativity, we now employ the cubic phase state to obtain concrete bounds.

Assume now we have for the Wigner negativity of both cubic phase states that
\begin{align}
    \mathcal{W}(\e^{3\Delta r} \e^{3r}\gamma)-\mathcal{W}(\e^{3r}\gamma)=2\epsilon
\end{align}
so that $\phi$ and $\psi$ satisfy \eqref{eq:Wigner negativity difference} and $\Delta r$ is a small difference in the squeezing of the cubic phase states.
As we have seen in the previous section the fidelity between two cubic phase states only depends on the difference in squeezing if the $\gamma$ is the same in both states.

To summarize, the fidelity depends only on the squeezing difference $\Delta r$, whereas the difference in Wigner negativity, and thus in $\epsilon$, is governed by its scaling with $x = e^{3r}\gamma$.
As we have shown in the previous section, the Wigner negativity scales faster than a power law with some $\alpha$.
Then $\epsilon$ is
\begin{align}
    2\epsilon =\qty(\e^{3\Delta r} \e^{3r}\gamma)^\alpha - \qty(\e^{3r}\gamma)^\alpha =\qty(\e^{\alpha \Delta r}-1) \qty(    \e^{r}\gamma)^\alpha
\end{align}
and therefore
\begin{align}
    \Delta r= \alpha \log\qty(\frac{2\epsilon}{\qty(\e^{3r}\gamma)^\alpha}+1).
\end{align}
We see that if we have a lower bound on the scaling of the Wigner negativity with $x$ then we have an upper bound on $\Delta r$ and a lower bound on the fidelity $F(\psi,\phi)$. 
Thus in the limit $x=\e^{3r}\gamma\rightarrow \infty$ we obtain $\e^{3\alpha r}\gamma^{\alpha} \rightarrow \infty$ and in consequence $\Delta r\rightarrow 0$.
This means that $F(\psi,\phi)\rightarrow 1$ and the number of samples $N(\epsilon, \delta)$ diverge for constant $\epsilon$ and $\delta$.
Since the Wigner negativity of the cubic phase state grows faster than the lower bound we derived, maintaining a constant $\epsilon$ requires $\Delta r$ to decrease more rapidly than in the case of the bound.
This means that generally the sampling lower bound diverges, even when only considering one mode.

Therefore, to make a meaningful statement, we must take the mean photon number into account. While keeping the difference in Wigner negativity constant, $\Delta r$ decreases as the Wigner negativity increases.

\begin{figure}
    \centering
    \includegraphics[width=0.6\linewidth]{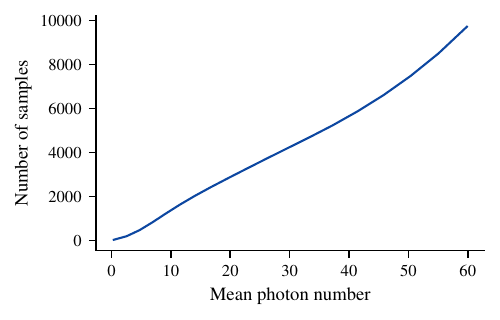}
    \caption{This figure shows the number of samples required over the mean photon number using directly the values for the Wigner negativity of a cubic phase state.}
    \label{fig:sample_direct}
\end{figure}

Let us put all the ingredients together and investigate how the number of samples scale with the mean photon number of the cubic phase state. 
In general, for fixed accuracy $\epsilon$, $\Delta r$ depends on the Wigner negativity $\mathcal{W}$, which itself is a function of the parameter $x$. This dependence allows us to relate $\mathcal{W}$ to the mean photon number of the state.
In order to compute the lower bound we fix $\epsilon, \delta$ and then compute Wigner negativity for increasing $x$ such that the difference is $2\epsilon$. Using these $x$ we compute the mean photon number. 
In Figure~\ref{fig:sample_direct}, we plot this lower bound on the required number of samples for estimating the Wigner negativity.

To recap, the fidelity depends on the difference in squeezing in two cubic phase states, whose difference in Wigner negativity is constant $2\epsilon$. 
A lower bound on the scaling of the Wigner negativity yields an upper bound on the squeezing difference, and consequently a lower bound on the bound we obtain by directly evaluating the winger negativity exactly.
To reveal how the number of required samples scales with $x$, we will evaluate the lower bound assuming the Winger negativity follows a power law scaling.
In the previous section we showed that the Wigner negativity scales for large $x$ at least with $x^{\frac{1}{6}}$, while we see numerically that it grows faster than $x^{\frac{1}{3}}$.
Then
\begin{align}
    \Delta r= \log\qty(\frac{\epsilon}{\mathcal{W}}+1).
\end{align}
We directly see how $\Delta r$ decreases by increasing the Wigner negativity $\mathcal{W}$. 
The sampling lower bound is then given as
\begin{align}
    N(\epsilon,\delta)\geq \frac{\log\qty(\frac{1}{4\delta(1-\delta)})}{\log(\cosh(\Delta r(\epsilon,x)))}&= \frac{\log\qty(\frac{1}{4\delta(1-\delta)})}{\log\qty(\frac{1}{2}\qty[ \frac{2\epsilon}{\mathcal{W}}+1+ 
    \frac{1}{\frac{2\epsilon}{\mathcal{W}}+1}]    )}\\
    &= \frac{\log\qty(\frac{1}{4\delta(1-\delta)})}{\log\qty(\frac{1}{2}\qty[ \frac{\epsilon}{x^{\alpha}}+1+ 
    \frac{1}{\frac{\epsilon}{x^{\alpha}}+1}]    )}.
\end{align}
We see that a lower bound of the Wigner negativity, especially a lower bound on the scaling with $x$, leads to a lower bound of the number of samples.
 
\end{document}